\documentclass[twocolumn,aps,amsmath,amssymb,prA]{revtex4}
\usepackage{bbm}
\usepackage{amsfonts} 
\usepackage{amssymb}
\usepackage[dvips]{graphicx}
\usepackage{epsfig}
\newcommand{\beq}{\begin{equation}}
\newcommand{\eeq}{\end{equation}}
\newcommand{\bee}{\begin{eqnarray}}
\newcommand{\eee}{\end{eqnarray}}
\newcommand{\been}{\begin{eqnarray*}}
\newcommand{\eeen}{\end{eqnarray*}}
\newcommand{\da}{\dagger}

\newcommand{\bmm}{\begin{matrix}}
\newcommand{\emm}{\end{matrix}}

\def\d{\mathrm{d}}
\begin{document}

\input epsf.sty

\title{Entanglement entropy in quantum spin chains with broken 
reflection symmetry}

\author{Zolt\'an K\'ad\'ar}\email{kadar@isi.it}
\author{Zolt\'an Zimbor\'as}\email{zoltan.zimboras@isi.it}
\affiliation{Institute for Scientific Interchange Foundation,\\Villa Gualino, 
Viale Settimio Severo 75, 10131 Torino, Italy}

\date{\today}

\begin{abstract}
We investigate the entanglement entropy of a block of $L$ sites
in quasifree translation-invariant spin chains concentrating on the 
effect of reflection symmetry breaking. The majorana 
two-point functions corresponding to the Jordan-Wigner transformed
fermionic modes are determined in the most general case; from these 
it follows that reflection symmetry in the ground state
can only be broken if the model is quantum critical. The 
large $L$ asymptotics of the entropy is calculated analytically
for general gauge-invariant models, which has, until now, 
been done only for the reflection symmetric sector. Analytical 
results are also derived for certain non-gauge-invariant models, 
e.g., for the Ising model with Dzyaloshinskii-Moriya interaction. 
We also study numerically finite chains of length $N$
with a non-reflection-symmetric Hamiltonian and report that 
the reflection symmetry of the entropy of the first $L$ spins
is violated but the reflection-symmetric Calabrese-Cardy formula
is recovered asymptotically. Furthermore, for non-critical 
reflection-symmetry-breaking Hamiltonians we find an anomaly 
in the behavior of the "saturation entropy" as we approach 
the critical line.   
The paper also provides a concise but extensive review of the
block entropy asymptotics in translation invariant quasifree
spin chains with an analysis of the nearest neighbor case and the
enumeration of the yet unsolved parts of the quasifree landscape.
\end{abstract}

\maketitle

%%%%%%%%%%%%%%%%%%%%%%%%%%%%%%%%%%%%%%%%%%%%%%%%%%%%%%%%%%%%%%%%%%%%%%%%
\section{Introduction}
%%%%%%%%%%%%%%%%%%%%%%%%%%%%%%%%%%%%%%%%%%%%%%%%%%%%%%%%%%%%%%%%%%%%%%%%

Understanding the entanglement properties of systems with many degrees
of freedom, such as quantum spin chains, has been one of the main recent 
research topics connecting quantum information theory and condensed matter 
physics \cite{VLRK,CC,ECP,AFOV,PE}. Huge amount of results have 
been accumulated about translation-invariant systems. However, the 
results almost exclusively  
correspond to reflection symmetric systems, 
despite the fact that models violating reflection invariance play 
a prominent role in many-body theory, e.g., in describing 
interactions of Dzyaloshinskii-Moriya type or non-equilibrium steady states.

Considering a subsystem $\mathcal{S}$ of a system, which is in a pure state, 
the entanglement between the subsystem and its environment is characterised by
the von Neumann entropy
\bee
S(\rho_{\mathcal{S}}):= - {\rm{Tr}} (\rho_{\mathcal{S}} \ln 
\rho_{\mathcal{S}})\ , \nonumber
\eee 
where $\rho_{\mathcal{S}}$ denotes the density matrix of the subsystem.
In the case of infinite one-dimensional critical chains,
this entanglement entropy belonging to a block of $L$ contiguous spins
was shown to grow asymptotically as \cite{VLRK,CC}
\bee
S_L=\frac{c}{3} \ln L + k\ , \label{entasymp}
\eee
where $c$ is the conformal charge of its universality class and $k$
is a non-universal constant. For non-critical chains the asymptotics
of the entanglement entropy is bounded. This saturation value of the
entropy diverges as one approaches the critical point: 
it increases as \cite{CC}
\bee
S_{sat}=\frac{c}{3} \ln \xi +k'\ , \label{noncrit}
\eee
where $\xi$ is the correlation length. 
In the case of finite chains (with open boundary conditions) consisting 
of $N$ spins, the conformal field theoretic 
prediction for the entanglement entropy of the first $L$ spins (at 
criticality) is \cite{CC,LSCAe,LSCA}
\bee
S(L,N)=
\frac{c}{6} \ln \left (\frac{2N}{\pi} \sin \frac{\pi L}{N} \right) + \ln g +
\frac{k}{2}\ , \label{finsize}
\eee
where $\ln g$ is the boundary entropy introduced by Affleck 
and Ludwig \cite{AL}.
%For non-critical systems the asymptotics of the entanglement 
%entropy is bounded. 

In this paper, we will study the asymptotics of the entanglement entropy
in chains with broken reflection symmetry. 
We consider quasifree models (with finite range coupling): 
their Hamiltonian can be 
mapped to quadratic fermionic chains by the
Jordan-Wigner transformation \footnote{
Throughout this paper we will use the following convention for the Jordan-Wigner transformation:\newline
$\sigma_j^x=\left(\prod_{l=1}^{j-1}(2 b_l\,b_l^\dagger-1)\right)(b_j+b_j^\dagger)$, \newline
$\sigma_j^y=\left(\prod_{l=1}^{j-1}(2 b_l\,b_l^\dagger-1)\right)i\,(b_j^\dagger-b_j)$ \newline
$\sigma_j^z=2\,b_j\,b_j^\dagger-1$
}
\beq
H=\sum_{i,j=1}^N \left(A_{i,j}^{\phantom*}b_i^{\dagger}b_j^{\phantom\dagger} +
\frac{1}{2}B_{i,j}^{\phantom*}b_{i}^{\dagger}b_{j}^{\dagger}-
\frac{1}{2}B_{i,j}^{*}b_i^{\phantom\dagger}
b_{j}^{\phantom\dagger}\right)\ . \label{Hqfree}
\eeq
Throughout the paper we will assume either open boundary conditions 
or "fermionic" periodic boundary conditions ($b_i=b_{i+N}$) \footnote{Note 
that periodic boundary conditions on the fermion chain may not be mapped
to periodic boundary condition after the Jordan-Wigner transformation,
as was shown in \cite{LSM}.}.
The requirement of translation-invariance implies that 
$A$ and $B$ are Toeplitz matrices 
($A_{i+n,j+n}=A_{i,j}$ and $B_{i+n,j+n}=B_{i,j}$ 
for any $n\in {\mathbb N}$), 
hermiticity of $H$ implies that $A$ 
is a (possibly complex) hermitian matrix, and $B$ 
is (a possibly complex) anti-symmetric matrix. 
Finite-ranged interaction means that there exists a positive integer $n_0$ 
such that
$A_{0,l}=B_{0,l}=0$ if $l \ge n_0$.
Such a spin-chain Hamiltonian is not invariant with respect to the
reflection transformation $R(\sigma^{a}_i)=\sigma^{a}_{-i}$ $(a=x,y,z)$, 
iff $A$ is not a real matrix. (One might think that the 
term $(b_ib_j - b_jb_i)$, with $i > j$ 
also breaks the translation invariance of the 
spin chain, but a short calculation shows that 
its image under the Jordan-Wigner transformation is the following 
reflection-invariant 
term $\sigma^{-}_{i}\Pi_{n=i+1}^{j-1} \sigma^{z}_{n} 
\sigma_{j}^{-} + \sigma^{-}_{j}\Pi_{n=j-1}^{i+1} \sigma^{z}_{n} 
\sigma_{i}^{-}$). 
One of the most studied quantum spin chain with 
broken reflection symmetry is the Ising model with transverse magnetic 
field and Dzyaloshinskii-Moriya (DM) interaction 
(in the $z$-direction) \cite{DerMo,DVKB,JKLS,LL}:
\bee
H=\sum_{i=1}^N \sigma^x_i \sigma^x_{i+1} + h \sigma^z_{i} + 
D(\sigma^x_{i} \sigma^y_{i+1}-\sigma^y_{i} \sigma^x_{i+1})\ ,
\eee
Another type of model that has been studied extensively in the literature is
the model
\begin{eqnarray}&&
{\displaystyle\!\!\!\!\!\!\!H=\sum_{i=1}^N\Big( J(\sigma^{x}_i\sigma^{x}_{i+1}+\sigma^{y}_{i}\sigma^{y}_{i+1})+}
\label{current}
\\&&{\displaystyle
\!\!\!\!\!\!\!h(\sigma^{z}_i+\sigma_{i}^{x}\sigma_{i+1}^{y}\!-
\sigma_{i}^{y}\sigma_{i+1}^{x})}
+ \, \lambda \, \sigma^{z}_i(\sigma_{i-1}^{y}\sigma_{i+1}^{x}
-\sigma^{x}_{i-1}\sigma_{i+1}^{y})\Big),\nonumber\end{eqnarray}
whose ground states are used to describe the energy current carrying
eigenstates of the $XX$ model \cite{ARRS,ARRS2,EZ}. Certain non-reflection
invariant quasifree states also appear as invariant states of
reflection-invariant quantum cellular automata \cite{GUWZ}.

The entanglement entropy asymptotics of the models given by 
Eq.~(\ref{Hqfree}) has been studied 
by many authors \cite{jinkorepin, mezzrand1,mezzrand2, itsmezzadrimo,
itsjinkorepin}.
The main analytic tool for tackling this problem was expressing the entropy
in terms of the determinant of a Toeplitz matrix, applied first by
Jin and Korepin \cite{jinkorepin}.
Until now the most general results have been achieved by 
Keating and Mezzadri \cite{mezzrand1}, who gave a general analytic 
expression for the entropy asymptotics when $A$ is real and $B\equiv 0$, 
and by Its, Mo, and Mezzadri \cite{itsmezzadrimo}, who gave an 
analytic (although less explicit) expression
even for the case of general (finite-ranged)
real  $A$ and $B$ matrices, while certain results
about the $d$ dimensional case can be found in \cite{multidim}. However, 
none of these studies concerned reflection symmetry breaking
cases, i.e., when $A$ is complex. 

We will generalize the above mention results by deriving 
an analytic expression for the general gauge-invariant 
case (i.e., when $A$ is a "general" complex Hermitian finite-ranged 
Toeplitz matrix, while $B\equiv 0$). This includes,
as a particular case, the model described in Eq.~(\ref{current}). 
Moreover, we will also introduce a multitude of transformations
between models of Eq.~(\ref{Hqfree}), which 
allows for deriving analytic expressions for cases
with non-vanishing $B$. A remarkable result that 
we obtained is that for these "quasifree" models
reflection invariance can only 
be broken in the ground state if the model is critical. If the
model is non-critical the ground state of the model does not change
if we replace $A_{i,j}$ with
Re($A_{i,j}$) in the Hamiltonian. From this, as we will show, 
it follows that scaling in Eq.~(\ref{noncrit}) may be violated.
 However, we will discuss how 
we can reinterpret this equation to keep its validity.
Furthermore, we will present numerical results in non-reflection-symmetric
spin chains providing an example of broken reflection symmetry in the finite
size scaling of the entropy $S(L,N)\ne S(N-L,N)$ breaking the 
symmetry of Eq.~(\ref{finsize}), but we will see that 
that this deviation goes to zero as we increase the 
system size. 

The paper is structured as follows. In Section II we calculate
the majorana two-point functions of 
these general (finite-ranged) quasifree models and recapitulate
how one can obtain the entanglement entropy from the two-point functions. 
The results already known about the entanglement
asymptotics of certain types of quasifree models are collected in Section III. 
We derive an analytic formula for the entanglement entropy
for general gauge-invariant models in Section IV, whereas in Section V
we show how we can extend our results for certain types of non-gauge-invariant
models too. Section VI is an application of the above to models with
nearest neighbor interactions, while in Section VII we discuss how 
some of our analytic and numerical results conflict with the formulas 
(\ref{noncrit}) and (\ref{finsize}) and
how we can "resolve" this discrepancy. Finally, Section VIII is devoted 
to the summary and the remaining open questions. 
%%%%%%%%%%%%%%%%%%%%%%%%%%%%%%%%%%%%%%%%%%%%%%%%%%%%%%%%%%%%%%%%%%%%%%%%
%\section{Entropy\label{ent}}
%%%%%%%%%%%%%%%%%%%%%%%%%%%%%%%%%%%%%%%%%%%%%%%%%%%%%%%%%%%%%%%%%%%%%%%%%

%%%%%%%%%%%%%%%%%%%%%%%%%%%%%%%%%%%%%%%%%%%%%%%%%%%%%%%%%%%%%%%%%%%%%%%%%
\section{Two-point function of the Majorana operators and 
entanglement entropy}
%%%%%%%%%%%%%%%%%%%%%%%%%%%%%%%%%%%%%%%%%%%%%%%%%%%%%%%%%%%%%%%%%%%%%%%%%%

The entanglement entropy asymptotics of the models described 
by the quadratic Hamiltonians in Eq.~(\ref{Hqfree})
can be calculated from the ground-state expectation 
values $\langle m_k m_l \rangle$, where $m_n$'s 
denote the so-called
majorana operators defined as
\beq m_{2n}=i(b_n-b_n^\dagger),\quad m_{2n-1}=b_n+b_n^\dagger\ . 
\label{major} \eeq
In this section we will first derive
these majorana two-point functions in terms of the
matrices $A$ and $B$ that define
the Hamiltonian Eq.~(\ref{Hqfree}). 
Then we describe how to calculate (in this quasifree setting)
the entanglement entropy alone from two-point functions, 
and finally we recapitulate the "determinant trick" of
Jin and Korepin, which will allow us later to obtain analytical
results.
%, i.e., we 
%present how one can obtain the entanglement entropy asymptotics 
%from the determinant of the majorana two-point matrix.  

\subsection{The majorana two-point functions \label{Twopoint}}

Let us fix our conventions used in the calculation. We will
consider the "fermionic" periodic boundary condition: $b_i=b_{i+N}$.
The Fourier and inverse 
transforms of the one-particle annihilation operators read
\bee \tilde{b}_k&=&\frac{1}{\sqrt{N}}
\sum_n\exp\left(-\frac{2\pi i n k}{N}\right) 
b_n\label{fourm}\\
     b_n&=&\frac{1}{\sqrt{N}}\sum_k\exp\left(\frac{2\pi i n k}{N}\right) 
\tilde{b}_k,
\eee
the summation runs in the set of integers 
$\left[\frac{N-1}{2},\frac{N-1}{2}\right]$ 
($\left[-\frac{N}{2},\frac{N}{2}-1\right]$) 
for N odd (even) and the transform of the one-particle
creation operators are to be computed by means 
of taking the adjoint of the above formulae.
For Toeplitz matrices we define the Fourier transform as
\bee
%\begin{array}{ccc}
     X_k&=&\sum_n\exp\left(-\frac{2\pi i n k}{N}\right)X_{0,n}\\
     X_{0,n}&=&\frac{1}{N}\sum_k\exp\left(\frac{2\pi i n k}{N}\right)X_k
%\end{array}
\eee
here $X_{0,n}$ stands for either $A_{0,n}$ or $B_{0,n}$, and
the summation again runs in the set of integers 
$\left[\frac{N-1}{2},\frac{N-1}{2}\right]$ 
($\left[-\frac{N}{2},\frac{N}{2}-1\right]$) 
for N odd (even).
Using these definitions, the Hamiltonian (\ref{Hqfree}) can be written as
\beq
H=\sum_k\left(A_k^{\phantom*}\tilde{b}_k^{\dagger}
\tilde{b}_k^{\phantom\dagger} +
\frac{1}{2}B_k^{\phantom*}\tilde{b}_k^{\dagger}\tilde{b}_{-k}^{\dagger}-
\frac{1}{2}B_k^{*}\tilde{b}_k^{\phantom\dagger}
\tilde{b}_{-k}^{\phantom\dagger}\right)\ .
\eeq
To bring this Hamiltonian into a diagonal form 
$H=\sum_k \Lambda_k\,c^\da_k c_k,\;\;(\Lambda_k\in{\mathbb R})$, 
one performs a Bogoliubov transformation 
\beq c_k=\alpha_k \tilde{b}_k+\beta_k \tilde{b}^\da_{-k}\quad\alpha_k,
\beta_k\in {\mathbb C}\ ,\label{bt}\eeq 
where the coefficients $\alpha_k,\beta_k$ have to satisfy
\bee \alpha_k\beta_{-k}+\beta_k\alpha_{-k}&=&0\\
     |\alpha_k|^2+|\beta_k|^2&=&1\ ,\eee    
so that the canonical anticommutation relations 
$\{c_k,c^\dagger_{k'}\}=\delta_{kk'}$ are satisfied. 
The consistency conditions for the commutator $[c_k,H]=\Lambda_k\,c_k$ give
\beq \left(\begin{array}{rl}-A_k&B_k^*\\B_k&A_{-k}\end{array}\right)
\left(\begin{array}{l}\alpha_k\\\beta_k\end{array}\right)
=\Lambda_k\left(\begin{array}{l}\alpha_k\\
\beta_k\end{array}\right) \label{ee}\ .\eeq
One readily extracts the one-particle spectrum
\beq \Lambda_k=\frac{A_{-k}-A_k+
\sqrt{(A_k+A_{-k})^2+4B_kB_k^*}}{2} \, ,
\label{ops}\eeq
having taken the relations $A^*_k=A_k,\;B_{-k}=-B_k$ (which are direct 
consequence of $A_{j,i}^*=A_{i,j}, 
B_{j,i}=-B_{i,j}$) into account.
The ground state correlations for the two-point functions of the new 
Fermi operators read 
\bee \langle c_k^{\dagger}\,c_{k'} \rangle &=& \frac{1}{2}
\left(-\frac{\Lambda_k}{|\Lambda_k|} +1 \right)\delta_{k,k'} 
\nonumber 
%\\
%\langle c_k\,c_{k'}^{\dagger} \rangle &=&
%\frac{1}{2}\left(\frac{\Lambda_k}{|\Lambda_k|} +1 \right) \delta_{k,k'}
%\nonumber
\eee
and all other correlations vanish. Now, using the 
inverse of 
(\ref{bt}), $\tilde{b}_k=\alpha^*_k c_k+\beta_{-k}\,{c^\da}_{-k}$, 
one can compute the correlations among the Fourier 
components $\tilde{b}_k,\,\tilde{b}_k^\da$, and substituting the
solution of (\ref{ee}) for $\alpha_k,\beta_k$ we arrive at
\begin{eqnarray}
\langle b_j\,b_{\,l} \rangle\!\!&&=\!\!\frac{1}{N}
\sum_k\exp\frac{2\pi ik(j-l)}{N}\,\frac{B_k}{2\sqrt{\Delta_k}}
\left(\frac{\Lambda_k}{|\Lambda_k|}\!+
\!\frac{\Lambda_{-k}}{|\Lambda_{-k}|}\right) \nonumber \\
\langle b^\da_j\,b_{\,l}\rangle\!\!&&=\!\!\frac{1}{N}
\sum_k\exp\frac{2\pi ik(j-l)}{N}\times \label{bbp} \\ &&
 \frac{2 +\!
\left(\frac{\Lambda_k}{|\Lambda_k|}\!-\!
\frac{\Lambda_{-k}}{|\Lambda_{-k}|}\right) 
\!+\!\frac{A_{-k}+A_k}{\sqrt{\Delta_k}}
\left(\frac{\Lambda_k}{|\Lambda_k|}\!+
\!\frac{\Lambda_{-k}}{|\Lambda_{-k}|}\right)
}{4}\, , \nonumber
\end{eqnarray}
where $\Delta_k=(A_k+A_{-k})^2+4B_kB_k^*$, and the two remaining
two-point functions $\langle b^{\dagger}_j b^{\dagger}_l \rangle$ and 
$\langle b^{\phantom\dagger}_j b^{\dagger}_l \rangle$ can be calculated
directly from the above equations.
Ultimately, we would like to have a linear combination of the above, 
the two point functions of the self-adjoint 
majorana operators defined in Eq.~(\ref{major}).
%
%\beq m_{2j}=i(b_j-b_j^\dagger),\quad m_{2j-1}=b_j+b_j^\dagger\ . 
%\label{major} \eeq
Before writing down the final result, let us introduce some notations. 
We will use the combinations
\bee A^s_k\!&=&\!A_{-k}+A_k  \; , \; \;   A^a_k=A_{-k}-A_k\,
,\\ 
B^s_k\!&=&\!B_k+B^*_k  \; , \; \;  i\,B^{\,a}_k=B_k-B^*_k
\eee
and the step functions
\beq
M_k\!=\!\frac{1}{2}\!\left(\frac{\Lambda_k}{|\Lambda_k|}
-\frac{\Lambda_{-k}}{|\Lambda_{-k}|}\right)\!\!,  
\; P_k\!=\!\frac{1}{2}\!\left(\frac{\Lambda_k}{|\Lambda_k|}
+\frac{\Lambda_{-k}}{|\Lambda_{-k}|}\right)\!.
\eeq
Note, that $A^s_k\,,\,A^a_k\,,\,B^s_k\,,\,B^{\,a}_k\in {\mathbb R}$ 
and $\Delta_k=(A^s_k)^2+(B^s_k)^2+(B^{\,a}_k)^2$.
We now take the thermodynamic limit ($N \to \infty$) 
and write the final result in a manner usually adopted in the 
literature 
\bee
\langle m_jm_l\rangle=\delta_{jl}+i C_{jl}\ , \label{Cdef}
\eee
where the matrix $C$ has the following structure
\beq 
C = 
\begin{pmatrix} 
\ddots & \vdots & \vdots & \vdots & \vdots &   \cr
\cdots &  \; \Pi_{0} & \; \; \; \Pi_{-1} & 
\; \; \; \Pi_{-2} & \; \; \; \Pi_{-3} & \cdots \cr
 \cdots &  \; \Pi_{1} & \;  \Pi_{0} & \; \; \;
\Pi_{-1} & \; \; \; \Pi_{-2} & \cdots \cr
                      \cdots &  \; \Pi_{2} & \; \Pi_{1} & \; \Pi_{0} & 
\; \; \; \Pi_{-1} & \cdots \cr
\cdots & \; \Pi_{3} & \; \Pi_{2} & \; \Pi_{1} & \; \Pi_{0} & \cdots \cr
 & \vdots  & \vdots  & \vdots & \vdots & \ddots 
\end{pmatrix}\ .
\eeq     
The $\Pi_l$'s are $2\times 2$ block entries that read
\bee 
\begin{array}{l}
\Pi_{l}=
\frac{1}{2\pi}\int\limits_{-\pi}^{\pi} d\theta\, 
e^{-il\theta} \times \\
\left(\begin{array}{cc}{\displaystyle iM(\theta)
-P(\theta)\frac{iB^s(\theta)}{\sqrt{\Delta(\theta)}}}&{\displaystyle 
P(\theta)\frac{A^s(\theta)-iB^a(\theta)}
{\sqrt{\Delta(\theta)}}}\\\\{\displaystyle 
P(\theta)\frac{-A^s(\theta)
-iB^a(\theta)}{\sqrt{\Delta(\theta)}}}&{\displaystyle iM(\theta)+P(\theta)
\frac{iB^s(\theta)}{\sqrt{\Delta(\theta)}}}
\end{array}\right)
\end{array}
\label{symbol}\eee
where $\theta=2\pi ik/N$, so all Fourier series become functions 
on the circle $[0,2\pi]$ in the limit.
This type of matrix $C$ is called block-Toeplitz and the 
matrix argument in  (\ref{symbol}) of the 
integral $\varphi:S^1\to M_2({\mathbb C})$ is called its {\em symbol}.

The $n$-point majorana function can be obtained from the
two-point functions by the Wick rule \cite{LSM}:
\bee
&\langle m_{i_1} \ldots m_{i_{2k-1}} \rangle =0 \ , \nonumber
\\ 
& \langle  m_{i_1} \ldots m_{i_{2k}} \rangle 
= \sum\limits_{\pi} {\rm{sgn}}(\pi) 
\prod\limits_{l=1}^{k} \langle m_{\pi(2l-1)} m_{\pi(2l)} \rangle\ ,\nonumber
\eee  
where the sum runs over all pairings of $\{1,2, \ldots, 2k\}$, i.e.,
over all permutations of the $2k$ elements which satisfy 
$\pi(2l-1) < \pi (2l)$ for $l\in \{1,2,\dots,k\}$ and $\pi(2l-1) < \pi (2l+1)$ for 
$l\in\{1,2,\dots k-1\}$.

Before coming to the calculation of the entropy, let 
us analyse the obtained result. 
The one particle spectrum (\ref{ops}) has the form of a sum of a reflection
invariant $\sqrt{\Delta(\theta)}/2$ and a non-invariant 
term $A^a(\theta)/2$ (Note that the real space reflection
$n\to -n$ corresponds to the Fourier space one 
$k \to -k$  as follows from the Fourier transform (\ref{fourm})).
The symbol (\ref{symbol})
characterizing the correlation matrix $\langle m_i m_j\rangle$ has 
a dependence on the 
non-reflection invariant part of the spectrum only via 
$M_\theta=(\Lambda(\theta)/|\Lambda(\theta)|
-\Lambda(-\theta)/|\Lambda(-\theta)|)/2$. This 
term, however, vanishes identically unless 
$\Lambda(\theta_0)=0$ at some $\theta_0\in [0,2\pi]$. 
In other words, non-critical quasifree
systems never break reflection invariance \footnote{Of course, various non-quasifree models exist with a gap
that break reflection symmetry, see e.g., \cite{g}.}. 
We will discuss some implications of 
this important fact in Sections \ref{Nearest} and 
\ref{Violations}.

\subsection{Calculation of  the entanglement entropy from the 
two-point functions \label{enttwopoint}}

\begin{figure}\includegraphics{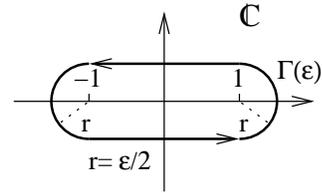}\caption{\label{contour}{The integration contour encloses a domain, 
where $e(1+\varepsilon,\lambda)$ defined in (\ref{fe}) is analytic and contains all singular values of 
the matrix $C_L$.}
} \end{figure}

Restricting the ground state to a subsystem consisting of
$L$ consecutive sites 
%(e.g., to the sites $1,2,3, \ldots , L$), 
one obtains a mixed state. Let us restrict the matrix $C$ defined in 
Eq.~(\ref{Cdef}) (which describes the two-point majorana 
correlations) to $L$ consecutive modes, that is to a $2L \times 2L$
submatrix
\beq 
C_L = 
\begin{pmatrix} \; \Pi_{0} & \; \; \; \Pi_{-1} & \cdots & \; \; \;
                \Pi_{-L+1} \cr
                 \; \Pi_{1} & \; \Pi_{0} & \cdots & \; \; \; \Pi_{-L+2} \cr
                      \vdots & \vdots & \ddots & \vdots \cr
                       \; \; \Pi_{L-1} &  \; \; \Pi_{L-2} & \cdots & 
                    \Pi_{0}
\end{pmatrix}
,
\eeq  
where the $\Pi_{l}$'s are $2 \times 2$ matrices given by Eq. (\ref{symbol}).
Let us denote by $W$ the orthogonal matrix, the adjoint action of which
brings the antisymmetric real matrix $C_L$ into its canonical form, i.e., 
for $(H_L)_{ij}=\sum_{k,l=0}^{2L-1}W_{ik}(C_L)_{kl}W_{jl}$, we have
\[
H_L= \bigotimes_{k=1}^{L} \nu_k 
\begin{pmatrix} 0 & 1 \cr
                                        -1 & 0  
\end{pmatrix},
\]
where $\nu_k \in [0,1]$ ($k=1,2, \ldots ,L$) are the singular 
values of $C_L$. (Due to the fact that $C_L$ is antisymmetric, 
the degeneracy of its singular values is always an even number, 
that is why we label them only from $1$ to $L$.)
The density matrix corresponding to the restricted state can be written
as \footnote{One can check that 
for any m 
$m_{i_1}m_{i_2}\cdots m_{i_{k}}$ ($1 \le i_1,i_2, \ldots i_k \le L$)
monomial of the majorana operators its
expectation value (given by the formulas ...) is equal to 
${\rm{Tr}}(\rho_L m_{i_1}m_{i_2}\cdots m_{i_{k}})$, hence $\rho_L$ is indeed
the density matrix of the restricted state.}
\[\begin{array}{l}
\rho_L= \\\\{\displaystyle\prod\limits_{j=1}^{L}\!\!\left[ \frac{1+\nu_j}{2}\cdot\frac{\mathbbm{1}
+i\hat{m}_{2j-1}\hat{m}_{2j}}{2}
+ \frac{1-\nu_j}{2}\cdot\frac{\mathbbm{1}-i\hat{m}_{2j-1}\hat{m}_{2j}}{2}\right]},
\end{array}\]
where $\hat{m}_j = \sum_{k=0}^{2L-1} W_{jk} m_k $ for all $j=0,1, \ldots 2L-1$.
(Actually, translation invariance is not used here, 
the density matrix of any quasifree state, i.e., of any state 
for which the Wick expansion applies, can be written in this form.)  
The entropy can now be easily calculated. It can be 
written in terms of the function
\bee
e(x ,\nu)\equiv - \frac{x+\nu}{2}\ln \frac{x+\nu}{2} - 
\frac{x-\nu}{2}\ln \frac{x-\nu}{2}\label{fe}
\eee
as 
\beq S_L\equiv S(\rho_L)=\sum_{j=1}^L e(1,\nu_j)\ \label{entgen}.\eeq
The trick \cite{jinkorepin} to obtain the asymptotics 
of the entanglement as the size of 
the block grows is computing the determinant
\beq
D_L(\lambda)=\det\,(i\lambda I+C_L)=
(-1)^L\prod_{j=0}^{L-1}(\lambda^2-\nu_i^2)\ ,
\eeq
and exploiting the residue theorem by writing down the following integral
\beq\lim_{\varepsilon\to 0}\frac{1}{4\pi i}\oint_{\Gamma(\varepsilon)} 
e(1+\varepsilon,\lambda)\frac{d\ln (D_L(\lambda)(-1)^L)}{d\lambda}\ ,
\label{contint}
\eeq
where the contour $\Gamma(\varepsilon)$ is shown in Fig.~\ref{contour}.  
That contour encircles all eigenvalues of 
$C_L$, but bounds a region, in which $e(1+\varepsilon,\lambda)$ is analytic. 
Hence, the main task in all cases is to compute the determinant of the
block-Toeplitz matrix matrix 
\bee
\widetilde{C}_L(\lambda) = i\lambda I+C_L \,  \label{Ctilde}
\eee
Finally, we should mention 
that in the gauge-invariant case (i.e., when $B=0$)
there is an easier method for the calculation
of the entanglement entropy. In this case, as can be seen from 
Eq.~(\ref{bbp}), 
the non-gauge-invariant two-point functions vanish 
($\langle b_j b_k \rangle=\langle b^{\dagger}_j b^{\dagger}_k \rangle =0$).
If we restrict the state to $L$ consecutive sites, and denote
by $M_L$ the corresponding restriction of the matrix
$M_{ij}=\langle b^{\dagger}_i b^{\phantom\dagger}_j \rangle$ and 
by $U_L$ the (not necessarily real) unitary, the adjoint action
of which diagonalizes $M_L$ 
($\sum_{j,k=1}^{L} U^*_{ij}(M_L)_{jk}\,U_{lk}=\lambda_i \delta_{il}$), 
then the density matrix of the restricted state reads 
\bee
\rho_L= \prod\limits_{i}^{L} \left(
\lambda_i c^{\dagger}_i c^{\phantom\dagger}_i + (1-\lambda_i)c^{\phantom\dagger}_i
c^{\dagger}_{i}  \right) \, , \nonumber
\eee
where $c_{i}=\sum_{j} U_{i,j} b_j$ and $\lambda_i\in [0,1]$.
Hence the entropy of the restricted state is given by 
\beq S_L= - {\rm Tr} \, \rho_L \ln \rho_L = - \sum_{i=1}^{L} \left( 
\lambda_i \ln \lambda_i +(1-\lambda_i) \ln (1-\lambda_i) \right)\ .\label{entgint}\eeq
\section{Summary of previously known cases}

In this section we shortly recapitulate what has been previously
known about the entanglement entropy for quasifree
models.
\subsection{Gauge and reflection invariance \label{gri}} 

In the case of gauge- and reflection-invariant quasifree models, the 
matrix $B$ is zero while $A$ is real, which implies
$B^{s} (\theta) \equiv B^{a}(\theta) \equiv A^a(\theta) \equiv 0$ and the 
symbol of the majorana two-point functions (\ref{symbol}) reduces to
\beq \varphi(\theta)=\left(\begin{array}{cc} 0&
\frac{A^s(\theta)}{|A^s(\theta)|}\\-\frac{A^s(\theta)}{|A^s(\theta)|}&0
\end{array}\right) \ . \nonumber
\eeq 
Hence, $C_L$ can be factorized as
\bee
C_L(\lambda)= \left(\begin{array}{cc} 0& 1
\\-1&0 
\end{array}\right) \otimes G_L(\lambda)\ ,
\eee
where $G_L(\lambda)$ is the restriction of the Toeplitz matrix 
with scalar symbol 
$g(\theta)=A^s(\theta)/|A^s(\theta)|$ 
to an $L \times L$ block on the diagonal. From this it follows that
$\det(\widetilde{C}_L)=D_L=(-1)^L \det (\lambda I +G_L) \det (\lambda I - G_L)$.
To extract the entropy asymptotics, one only needs to calculate 
the ($L \to \infty$ asymptotics of the) determinant of 
$(\lambda I \pm  G_L)$ using the
Fisher-Hartwig theorem, and then use the residue theorem 
as described in Section \ref{enttwopoint}. This was done by 
Keating and Mezzadri \cite{mezzrand1,mezzrand2}: they
obtained the following result: Let there be number $R/2$ 
zeros of $A^s(\theta)$ denoted by $\theta_r$ ($r=1,\ldots ,R/2$)
in the semi-circle $[0, \pi]$ (implying another $R/2$ 
zeros in the other semi-circle $[-\pi,0]$: $-\theta_r$, $r=1, \ldots ,R/2$). 
Then the entanglement entropy asymptotics is 
given by
\beq S_L(\rho_A)=\frac{R}{6}\ln L+\frac{R}{6}K-\frac{R}{2}(\ln 2)\,I_3\ ,\eeq
where 
\bee K&=&1+\gamma_E+\frac{1}{R}\sum_{r=1}^{R/2} \ln |1-e^{2i\theta_r}|- 
\nonumber\\
&&\frac{2}{R}\sum_{1\leq s\leq r\leq R/2}(-1)^{r+s}\ln
\left|\frac{1-e^{i(\theta_r-\theta_s)}}{1-e^{i(\theta_r+\theta_s)}}\right| 
 \, ,\nonumber
\eee
where $\gamma_E = ...$ is Euler's constant, and $I_3=0.0221603...$, independent of
$A(\theta)$ (consult \cite{jinkorepin} 
for its derivation).  
%Without going into the details, one immediately notices that the 
%logarithmic divergence
%appears if the equation $A(\theta)=0$ has solution(s), 
%(or when $g(\theta)$ has jumps): they are given by 
%$\pm\theta_r\in [0,\pm\pi],\,r=1,2,\dots,R/2)$.
%%%%%%%%%%%%%%%%%%%%%%%%%%%%%%%%%%%%%%%%%%%%%%%%%%%%%%%%%%%%%%
\subsection{Reflection invariance and real $B_{ij}$ \label{42}}

The other case that has already been discussed in the literature
is the case when both matrix $A$ and $B$ are real, i.e., when  
$A^a(\theta)\equiv B^s(\theta)\equiv 0$. In this case, the symbol reads 
\beq \varphi (\theta)=
\left(\begin{array}{cc} 0&
\frac{A^s(\theta)-iB^a(\theta)}{|A^s(\theta)-iB^a(\theta)|}\\
-\frac{A^s(\theta)+iB^a(\theta)}{|A^s(\theta)+iB^a(\theta)|}&0
\end{array}\right)\eeq
Here the idea, invented for the XY model in \cite{itsjinkorepin} 
and generalized for the present case in \cite{itsmezzadrimo},   
is to extend the domain of $\varphi:S^1\to M_2({\mathbb C})$ to 
the complex plane and use a 
theorem of Widom \cite{widom55}, which yields a formula 
of the block Toeplitz determinant at hand 
expressed in terms of Wiener-Hopf factors of the 
symbol: $\varphi(z)=U_+(z)U_-(z)=V_-(z)V_+(z)$, where
the matrices $U_+,V_+$ $(U_-,V_-)$ are analytic 
inside (outside) the unit circle. The factorization 
resides on the fact, that due to the assumption of finite range 
interaction, the functions
${\cal A}(z)={\cal A}(\exp(-i\theta))\equiv A(\theta)$, and 
${\cal B}(z)={\cal B}(\exp(-i\theta))\equiv B(\theta)$ 
are Laurent polynomials. One writes 
\beq \frac{{\cal A}^a(z)-{\cal B}^s(z)}{|{\cal A}(z)-{\cal B}(z)|}\equiv \frac{q(z)}{|q(z)|}=
\sqrt{\frac{q(z)}{q(1/z)}}=
\sqrt{\prod_{j=1}^{2n_0}\frac{z-z_j}{1-z_jz}} \nonumber
\eeq
with $z_i$ being the roots of the polynomial $p(z)=z^{n_0}q(z)$, 
where $n_0$ is the range of the coupling
(defined after (\ref{Hqfree})). Note, that the equality in the middle 
is a choice of analytic continuation as 
$q(z)^*=q(1/z)$ holds on the unit circle (as is obvious from the general 
form $q(\exp -i\theta)\equiv A^s(\theta)-iB^a(\theta)$. 
The non-analytic behaviour of the above
rational function is then the only thing that 
has to be taken care of and the factorization is done with
the help of theta functions living on the hyperelliptic 
surface of genus $n_0$ given by
\beq w^2=\prod_{j=1}^{2n_0}(z-z_i)(1-zz_i)\ .\label{surf}\eeq
The $XY$-model has 
$n_0=1$ thus the underlying Riemann surface is a torus,
while for for general finite ranged couplings $q(z)$ can be any degree $n_0$ 
Laurent polynomial, which satisfies $q(z)^*=q(1/z)$ on the unit circle.
The result (Theorem 3. of \cite{itsmezzadrimo}) for the logarithmic 
derivative reads
\beq\begin{array}{l}\frac{d\ln D_L(\lambda)}{d\lambda}\approx-
\frac{2\lambda L}{1-\lambda^2}+ \\\\
\frac{1}{2\pi}\oint\mbox{tr}
\left[\left(\frac{dU_+(z)}{dz}U^{-1}_+(z)+V_+^{-1}(z)
\frac{dV_+(z)}{dz}\right)G^{-1}(z)\right]dz \end{array}\label{mezzdet}\eeq
and the difference (rhs.$-$lhs.)$<C\rho^{-L}$ where the constant $\rho$ 
satisfies $1<\rho<\min\{|\lambda_i|:|\lambda_i|>1\}$
(the complex numbers $\lambda_i$ are the roots of $p(z)$ and 
their reciprocals). The saturation entropy is given by
\[
  S(\rho_A)=\frac{1}{2}
  \int_{1}^{\infty}\ln{{\Theta\left(\beta(\lambda)\overrightarrow{e}
        +{\tau\over 2}\right)\Theta\left(\beta(\lambda)
        \overrightarrow{e}-{\tau\over 2}\right)}
        \over{\Theta^2\left({\tau\over 2}\right)}}\d\lambda\ .
\]
This formula depends on the surface (\ref{surf}) via the theta functions (which are uniquely defined by
some quasi-periodicity properties along non-contractible curves on the surface); their definition 
and that of their arguments will be omitted here (see \cite{itsmezzadrimo}). We only remark that it 
is exactly at criticality, 
when the above surface becomes degenerate and the formula diverges.

%%%%%%%%%%%%%%%%%%%%%%%%%%%%%%%%%%%%%%%%%%%%%%%%%%%%%%%%%%%%%%%%%%%%%%%%%%%%%%%%%%%%%%%%%%%%%%%%%%%%
\section{Gauge invariant models in general} \label{gim}
%%%%%%%%%%%%%%%%%%%%%%%%%%%%%%%%%%%%%%%%%%%%%%%%%%%%%%%%%%%%%%%%%%%%%%%%%%%%%%%%%%%%%%%%%%%%%%%%%%%

The reason why one could give an explicit formula for the
entropy asymptotics in the reflection and gauge invariant 
case (when $A^a(\theta) \equiv B^a(\theta) \equiv B^s(\theta)=0$) 
and a less explicit one in the case when 
$A^a(\theta) \equiv B^s(\theta) \equiv 0$
was that the structure of the symbol $\varphi(\theta)$ 
was considerably simplified in both cases.
  
In the general quasifree case it is hard to find the Wiener-Hopf 
factorization of the symbol, since there is no identically zero 
entry of in the matrix function $\varphi (\theta)$.
This is true even in the restricted case of gauge invariant 
(but not reflection invariant) models. However, as we will show 
in this section, one can circumvent this problem in this restricted case.
We have seen in section \ref{enttwopoint} that 
we can extract the entropy also from the correlation matrix
$C'_L\equiv\langle b_ib^\dagger_j\rangle|_{i,j=1..L}$: it is given 
by $S_L=- \sum_{i=1}^{L} \left( \lambda_i \ln \lambda_i +(1-\lambda_i) 
\ln (1-\lambda_i) \right)$, where $\lambda_i$ are the 
eigenvalues of the matrix $C'_L$. 
Now, we can use the contour integral trick again with a
small alteration and write 
the entropy as
% Now, one can use the contour integral trick again with a small
% alteration. Although, as we will see shortly, this is not 
% a great mathematical departure from the previously existing
% results, it is crucial for what follows.
% It will be used again and again when we transform 
% more general non-gauge-invariant models to 
% generic gauge-invariant ones to obtain the entropy
% asymptotics. Now, getting back to the calculation,
% we can write the entropy as 
\bee
S_L= \lim_{\varepsilon\to 0}\frac{1}{2\pi i}\oint_{\Gamma(\varepsilon)} 
e(1+\epsilon,\lambda)\frac{d \ln D'_L(\lambda)}{d\lambda}\, ,
\label{cint2}
\eee
where $D'_L (\lambda)=\det\widetilde{C}'_L(\lambda)=
\det (\lambda I - (2C'_L-I))$, the function $e(x, \lambda)$ and the contour $\Gamma$
were defined in Section \ref{enttwopoint}. Hence the situation is analogous to 
section \ref{gri} except that $\lambda I-G_L$ is replaced by 
$\lambda I-(2C'_L-I)$, which is also a Toeplitz matrix, but its 
symbol 
\beq \lambda+1-2c'(\theta)=\lambda+\frac{A(\theta)}{|A(\theta)|}\label{ts}\eeq
 is not necessarily
symmetric ($A_{i,j}\neq A_{j,i}$ implies $c'(\theta)\neq c'(-\theta)$). 
Now we can use the 
{\em Fisher-Hartwig conjecture} \cite{fh}:
Suppose that the symbol $p(\theta): S^1 \to \mathbb{C}$ of a Toeplitz matrix
 has the following form 
\begin{equation}
p(\theta)=\psi(\theta) \prod_{r=1}^R t_{\beta_r,\,
\theta_r}(\theta) u_{\alpha_r,\,\theta_r}(\theta)
\label{fh}\end{equation}
with
\begin{eqnarray}
t_{\beta_r,\,\theta_r}(\theta)\!&=&\!e^{-i\beta_r (\pi-\theta+\theta_r)}, \; \theta_r<\theta <2\pi+\theta_r \nonumber\ ,\\
u_{\alpha_r,\,\theta_r}(\theta)\!&=&\!\Bigl(2-2\cos
(\theta-\theta_r)\Bigr)^{\alpha_r}, \; \mbox{Re}
\alpha_r>-\frac{1}{2}\ , \nonumber
\end{eqnarray}
where the function $\psi:S^1\to \mathbb{C}$ is smooth, non-vanishing and has zero winding number. Then 
the $\mathrm{L}\to \infty$ asymptotic formula 
for the determinant reads
\begin{eqnarray*}
\det P_{L}&=&
\left(\exp \left(\frac{1}{2\pi} \int_0^{2\pi}\ln
\psi(\theta) \mathrm{d} \theta\right)\right)^{\mathrm{L}} \times \nonumber\\
&& \left(\prod_{i=1}^R {\mathrm{L}}^{\alpha_i^2-\beta_i^2}\right)  
 {\cal E}[\psi, \{\alpha_i\}, \{\beta_i\},\{\theta_i\}]\ ,
\end{eqnarray*}
where
\begin{eqnarray}
&& {\cal E}[\psi, 
\{\alpha_i\},\{\beta_i\},\{\theta_i\}]\!=\!{\cal E}[\psi] \times  \nonumber\\
&&\times \prod_{i=1}^R G_B(1+\alpha_i+\beta_i) 
G_B(1+\alpha_i-\beta_i)/G_B(1+2\alpha_i)
\nonumber\\
&&\times \prod_{i=1}^R \biggl(\psi_-\Bigl(\exp(\mathrm{i}
\theta_i)\Bigr)\biggr)^{-\alpha_i-\beta_i} \biggl(\psi_+
\Bigl(\exp(- \mathrm{i} \theta_i)\Bigr)\biggr)^{-\alpha_i+\beta_i}\nonumber\\
&&\times \hspace{-0.4cm}\prod_{1\leq i \neq j \leq R}
\biggl(1-\exp\Bigl(\mathrm{i}
(\theta_i-\theta_j)\Bigr)\biggr)^{-(\alpha_i+\beta_i)(\alpha_j-
\beta_j)} \, , \nonumber
\end{eqnarray}
and the so-called Barnes function is defined by
\begin{eqnarray}
G_B(1+z)=(2\pi)^{z/2} e^{-(z+1)z/2-\gamma_E z^2/2} \times \nonumber \\
\prod_{n=1}^{\infty} \{ (1+z/n)^n e^{-z+z^2/(2n)}\} \, . \nonumber
\end{eqnarray}

In our case the symbol, defined by Eq.~(\ref{ts}) above, is a step function jumping between $\lambda+1$ 
and $\lambda-1$, and the jumps occur at the zeros of $A(\theta)$. 
 We can assume that $A(0)>0$, 
as the local transformation $\hat{b}_i=b^{\dagger}_i$ (which keeps the entanglement entropy invariant) 
yields $A(0)\to -A(0).$  
Using the notation for the zeros of $A(\theta)$ by $\theta_r$, $r=1,2,\dots R$ 
in an increasing order in the period $(0,2\pi]$ we can write the 
factors in (\ref{fh}) for the symbol (\ref{ts})
\bee
&&u_{{\alpha_r},\theta_r} = 1 \, , \nonumber\\
&&\Psi(\theta)=(\lambda+1)
\left(\frac{\lambda+1}{\lambda-1}\right)^{-\frac{1}{2\pi}\sum_j^{R/2} 
(\theta_{2j+1}-\theta_{2j})-1} \nonumber\\
&& t_{\beta_r,\,\theta_r}(\theta)\!=\!e^{-i\beta_r (\pi-\theta+\theta_r)},\quad
\; \theta_r<\theta <2\pi+\theta_r \, ,\nonumber 
\eee
where
\[\beta_r=(-1)^r\frac{1}{2\pi i}\ln\frac{\lambda+1}{\lambda-1}\ .\]

Indeed, one can easily check that the 
function given by (\ref{fh}) with the above defined
ingredients has the value $p(0)=\lambda-1$ and 
alternates between $\lambda\pm 1$ with jumps at the zeros of 
$A(\theta)$. Now, substituting our data in the statement 
of the conjecture we get the expression for the determinant
\[\begin{array}{l} {\displaystyle\det D'_L(\lambda)=(\ln\Psi)^L 
L^{-R \beta^2}
\displaystyle \prod_{{r\neq s}\atop 
{r=s\,mod\,2}}\left(1-e^{i(\theta_s-\theta_r)}\right)^{\beta^2}}\\\\
{\times \displaystyle
\prod_{{r\neq s}\atop
{r\neq s\,mod\,2}}\left(1-e^{i(\theta_s-\theta_r)}\right)^{-\beta^2}
\left(G_B(1+\beta)G_B(1-\beta)\right)^R}\ .\end{array}
\]
From this point, the calculation of the contour integral
(\ref{cint2}) is entirely identical to that of 
\cite{jinkorepin,mezzrand1}, and the result for
entropy asymptotics reads

\beq \begin{array}{l}S_L={\displaystyle\frac{R}{6}\ln L-\frac{1}{6}
\sum_{{r\neq s}\atop
{r=s\,mod\,2}}\ln(1-e^{i(\theta_s-\theta_r)})
+}\\\\
{\displaystyle \frac{1}{6}\!\!\sum_{{r\neq s}\atop
{r\neq s\,mod\,2}}\ln(1-e^{i(\theta_s-\theta_r)})+\frac{R}{6}\Big( (1+\gamma_E)-6I_3\ln 2\Big)}\ ,\end{array}
\label{entnob}\eeq
where the constants $\Gamma_E$ and $I_3$ were given at the end of 
section \ref{gri}. 
%%%%%%%%%%%%%%%%%%%%%%%%%%%%%%%%%%%%%%%%%%%%%%%%%%%%%%%%%%%%%%%%%%%%%%%%%
%%%%%%%%%%%%%%%%%%%%%%%%%
\section{Exact results for the entropy asymptotics for certain
non-gauge-invariant models} 
%%%%%%%%%%%%%%%%%%%%%%%%%%%%%%%%%%%%%%%%%%%%%%%%%%%%%%%%%%%%%%%%%%%%%%%
%%%%%%%%%%%%%%%%%%%%%%%%%%%
We now turn to discuss the cases of some non-gauge-invariant models. 
In the first two subsections we will determine the entropy asymptotics
for chains that are Kramers-Wannier selfdual and for those that
decouple to two independent majorana chains, by relating these cases
to certain gauge-invariant models. In the last two subsections we will relate the 
entropy asymptotics of different non-gauge-invariant models, by generalizing 
the XY-Ising transformation and doing local rotations. 

We will make use of the fact that one can write 
the general (translation-invariant) 
quasifree Hamiltonian (\ref{Hqfree}) in terms of the 
majorana operators defined by (\ref{major}) in the following way:
\beq H=i\sum_{j,l=1}^{2N} T_{j,\,l}\,m_jm_l \label{majorham} \eeq
with the properties $T_{j,\,l}=-T_{l,j}\in {\mathbb R}$ 
and $T_{j+2n,\,l+2n}=T_{j,\,l}$ for all $n\in {\mathbb Z}$. 
The transformation between the two descriptions reads
\been T_{2j-1,\,2l-1}&=&\frac{1}{4}\mbox{Im}(A_{j,\,l}+B_{j,\,l})\\
T_{2j,\,2l}&=&\frac{1}{4}\mbox{Im}(A_{j,\,l}-B_{j,\,l})\\
T_{2j-1,\,2l}&=&\frac{1}{4}\mbox{Re}(-A_{j,\,l}+B_{j,\,l})\\
T_{2j,\,2l-1}&=&\frac{1}{4}\mbox{Re}(A_{j,\,l}-B_{j,\,l})\ .
\eeen
%
%We can now enumerate the class of models that we can compute the 
%entropy for and the transformations, which lead to the
%solution. 

\subsection{Kramers-Wannier selfdual models}

The Kramers-Wannier (or disorder)
spin operators on a spin chain are
defined in terms of the original spin operators (Pauli matrices) as
\bee
\widehat{\sigma}_{l}^{x}=\prod_{i=1}^{l} \sigma_i^z \, , \; 
\widehat{\sigma}_{l}^{z}=\sigma_{l}^x \sigma_{l+1}^x \, ,\; 
\widehat{\sigma}_{l}^{y}=-i \widehat{\sigma}_{l}^x
\widehat{\sigma}_{l}^{z} \ . 
\eee
These spin operators also satisfy the Pauli commutation 
relations $[\widehat{\sigma}_l^a, \widehat{\sigma}_k^b]=
i \sum_c \delta_{k l}\epsilon_{a b c}\,\sigma_l^{c}$. If the Hamiltonian is
invariant with respect to the above transformation 
then it is said to be {\it Kramers-Wannier selfdual}. 
%with respect to the Kramers-Wannier duality. 
Such selfdual Hamiltonians always describe
critical models, an example is the critical point of the Ising model. 
A straightforward calculation shows that a
quasifree Hamiltonian (\ref{majorham}) is 
selfdual iff $T_{j,l}=T_{j+1,l+1}$ (recall, that translation invariance
only implies $T_{j,l}=T_{j+2,l+2}$). 
%This condition holds if
%
%\bee A_{l}&=&4i T_{2l}-2T_{2l-1}+2T_{2l+1}\\B_{l}&=&2T_{2l-1}+2T_{2l+1}\ ,\eee
%
%where the notation $X_l\equiv X_{0,l}$ was introduced for all the 
%quantities in the formula. 
Or in other words, the selfdual models is
the class of quasifree models, whose $B$ matrix is real and the 
equality Re$(A_{i,j}+B_{i,j})=$Re$(-A_{i,j+1}+B_{i,j+1})$ is satisfied 
for every integers $i,j$. The two-point functions are then given by:
\beq \langle m_j m_l \rangle=\delta_{jl}+
\frac{1}{2\pi}\int_{-\pi}^\pi e^{-i(j-l)\,\theta}
\frac{i\,T(\theta)}{|T(\theta)|}d\theta
\label{iglo}\eeq
with $T_\theta=\sum_n e^{-i\theta n}T_{0,n}$, thus in this case
the block-Toeplitz matrix  of the majorana expectation values
reduces to an ordinary Toeplitz matrix. 
Moreover, the above formula is (up to a factor of $2$) identical to that
of $\langle b^\dagger_jb_l\rangle$ for a gauge-invariant model 
with the symbol given as $A(\theta)=i T(-\theta)$.

Now, let us compare the calculations of the entropy from the matrix $\langle m_i m_j\rangle$ in the general quasifree case 
(cf. Eq.~(\ref{entgen})) and that from the matrix $\langle b^\dagger _i b_j\rangle$ in the 
gauge-invariant case Eq.~(\ref{entgint}). One can 
immediately conclude that the entanglement entropy of $L$ spins 
in a Kramers-Wannier selfdual model defined by
a matrix $T_{i,j}$ equals half the entanglement entropy of $2L$ spins 
in a gauge-invariant quasifree model defined
by $A_{i,j}=iT_{i,j}$. Thus our result in Section \ref{gim} applies to all 
Kramers-Wannier selfdual models as well. 
As an important example, we will apply this procedure in Section \ref{Nearest}
to obtain the analytic form of the entropy 
asymptotics for the critical Ising model with DM interaction.

\subsection{Directly decoupled majorana chains}

Next we turn to the case when the fermion chain
decouples to two separate majorana chains. 
From the form (\ref{majorham}) of the Hamiltonian we 
immediately see that if
$T_{2i,2j+1}=T_{2i+1,2j}=0$ for all $i,j$, which is equivalent 
to having purely imaginary matrices $A$ and $B$, 
then the fermion chain decouples to two independent majorana
chains: the one consisting of the odd modes and that of 
the even ones. Computing the symbol (\ref{symbol}) 
corresponding to $B^a\equiv A^s\equiv 0$ gives
\beq \varphi(\theta)=-i\left(\begin{array}{cc} 
\frac{-A^a(\theta)-B^s(\theta)}{|-A^a(\theta)-B^s(\theta)|}&0\\0&
\frac{-A^a(\theta)+B^s(\theta)}{|-A^a(\theta)+B^s(\theta)|}
\end{array}\right)\ ,\eeq
hence the matrix $C_L$ 
is a direct sum of two Toeplitz matrices with symbols 
$(-A^a \pm B^s)/|- A^a \pm B^s|$ corresponding to the
two uncoupled majorana chains. As in the previous section,
one can again relate the majorana expectation
values to the ground state expectation values of  
the $b_i b_j^{\dagger}$ 
operators of gauge-invariant models. Namely, we have
\bee
\langle b^\dagger_j b_l \rangle_{-A+B}&=&\frac{1}{2}\langle m_{2j} m_{2l}\rangle \\
\langle b^\dagger_j b_l\rangle_{-A-B}&=&\frac{1}{2}\langle m_{2j+1} m_{2l+1}\rangle
\eee
where $\langle\cdot\rangle_{-A\pm B}$ stand for the expectation values in the gauge invariant models with 
$H=\sum_{jl}(-A_{jl}\pm B_{jl}) b^\dagger_j b_l$. 
Thus, by virtue of Equations~(\ref{entgen}) and (\ref{entgint}) the entropy in the original model is given by the sum
of entropies in the gauge invariant ones above.
\subsection{The generalized XY-Ising correspondence}
There is also a less direct way certain fermion chains can be decoupled 
into two independent chains. Suppose that matrix $T$ in 
Eq. (\ref{majorham}) satisfies the following properties (for all $i,j$)
\bee
T_{4i,4j-1}= T_{4i,4j-2}=T_{4i-1,4j-3}=0 \, . \nonumber
\eee
By defining 
\bee
&& m^{(1)}_{2i-1}=m^{(1)}_{4i-3} \, , \; m^{(1)}_{2i}=m^{(1)}_{4i} \, , 
\nonumber\\
&&  m^{(2)}_{2i-1}=m^{(2)}_{4i-2} \, , \; m^{(2)}_{2i}=m^{(2)}_{4i-1} \, ,
\label{newmajor}\eee
one can see that the original quasifree Hamiltonian with $2N$ sites
can be written as the sum of two other quasifree Hamiltonians with
$N$ sites:
\bee
H&=& \sum_{i,j=1}^{4N} T_{i,j}m_im_j= \nonumber \\
&& \sum_{i,j=1}^{2N} T^{(1)}_{i,j}m^{(1)}_im^{(1)}_j + 
\sum_{i,j}^{2N}T^{(2)}_{i,j}m^{(2)}_im^{(2)}_j\, . \, \nonumber
\eee
Here the components of matrices $T^{(1)}$ and $T^{(2)}$ can be straightforwardly
matched with the components of matrix $T$ using the 
correspondence (\ref{newmajor}); it turns out that 
the decoupled subchains are also translation invariant:
$T^{(1)/(2)}_{i,j}=T^{(1)/(2)}_{i+2,j+2}$. This type of decoupling is the 
generalization of the famous XY-Ising correspondence \cite{XY-Ising} (for an
other type of recent generalization of this correspondence, see \cite{CR}).

Considering the ground state in the thermodynamic limit, this
type of decoupling immediately implies
that the entanglement entropy of $2L$ consecutive spins in the
model defined by $T$ equals the sum of the entropies of $L$ spins
for the models defined by $T^{(1)}$ and $T^{(2)}$. 
This method was used for deriving the entropy asymptotics of the critical
Ising model from that of the critical XY chain \cite{IJ},
our result generalizes this.

\subsection{On general reflection-invariant models}

As we have discussed, even for reflection-invariant chains 
($A_{i,j}$ real, $B_{i,j}$ complex) there is no general formula
for the entanglement entropy asymptotics. However, as we mentioned in 
Section \ref{42}, there is a formula for the saturation entropy in case 
the matrix $B$ is real. In this subsection we show that a 
subclass of models with 
complex $B$ can be
transformed back to the real case. 

A transformation 
on the vector $(m_{2j-1},m_{2j})\equiv v^j\mapsto Uv^j$ with a 
constant matrix $U\in U(2)$ is called canonical if 
the anticommutation relations 
$\{m_j,m_l\}=2\delta_{jl}$ are preserved. 
For the two point functions it results in the adjoint 
action $\langle v^j\otimes v^j\rangle\mapsto 
U\langle v^j\otimes v^j\rangle U^\dagger$. Assume now that there 
are constants $c_x,c_y,c_z\in {\mathbb R}$ with at least one of them 
non-vanishing, 
such that 
$c_x B^a(\theta)-c_y A^s(\theta)+c_z B^s(\theta)\equiv 0$ for 
all $\theta\in [0,2\pi)$. 
In this case there are rotations, which rotate the vector
$c\equiv (c_x,c_y,c_z)$ into $c'\equiv (0,0,c'_z)$ ($c'^z\neq 0$) 
and consequently $v\equiv(B^a,-A^s,B^s)$ into 
$v'\equiv(B'^a,-A'^s,0)$ (as $v\perp c\Rightarrow v'\perp c'$). 
Furthermore, the Toeplitz symbol can be written 
as $\varphi=iM\,{\bf 1}-iP/\sqrt{\Delta}\,(\sum_a\sigma^a v_a), 
(a=x,y,z)$, so the rotation can be done by the adjoint action 
of $SU(2)$ on $2 \times 2$ traceless Hermitian matrices
$U_R\,G(v)\,U_R^{-1}=G(Rv)$ ($G(v)\equiv G$). This is exactly 
the above defined local transformation. 
Note, that the invariance of the entropy can be 
immediately seen from the formula (\ref{mezzdet}), which is 
invariant under the simultaneous transformation of all 
matrices by the adjoint action of any constant matrix 
(and the Wiener-Hopf factorization remains also valid).
The general case, when $A^a,B^a,B^s$ are linearly independent 
Laurent polynomials of $e^{i\theta}$, 
this method does not
work. One could in principle try to follow 
a strategy similar to that of \cite{itsjinkorepin} as was 
done in \cite{itsmezzadrimo} sketched in section \ref{42}. 
To obtain explicit results, where physical limits can be studied,
is difficult.

%%%%%%%%%%%%%%%%%%%%%%%%%%%%%%%%%%%%%%%%%%%%%%%%%%%%%%%%%%%%%%%%%%%%%%%%%%%%%%%%%%%%%%%%%%%%%%%%%
\section{Nearest neighbor coupling \label{Nearest}}
%%%%%%%%%%%%%%%%%%%%%%%%%%%%%%%%%%%%%%%%%%%%%%%%%%%%%%%%%%%%%%%%%%%%%%%%%%%%%%%%%%%%%%%%%%%%%%%%%%

We will now look at the general quasifree model with 
nearest neighbour coupling and apply the 
above machinery to study its entanglement entropy. Our method yields analytic expression for
the Ising model with DM interaction at the critical point, while for the general 
non-critical case we demonstrate that the ground state is not effected by the DM term, hence the results
\cite{itsmezzadrimo} apply.

The Hamiltonian of the most general 
nearest neighbour spin chain that can be mapped
to a quasi-free fermion chain is given by \footnote{Note, that the term $\sigma_j^x\sigma_{j+1}^y+\sigma_j^y\sigma_{j+1}^x$ can be eliminated by the
basis change \newline $(\sigma_j^x,\sigma_j^y)\to((\sigma_j^x-\sigma_j^y)/\sqrt{2},(\sigma_j^x+\sigma_j^y)/\sqrt{2})$}
\bee H&=&\sum_j ((1+\gamma)\sigma_j^x\sigma_{j+1}^x+
(1-\gamma)\sigma_j^y\sigma_{j+1}^y+ \nonumber \\
&&D(\sigma_j^x\sigma_{j+1}^y-\sigma_j^y\sigma_{j+1}^x)+h\sigma_j^z)\ . 
\label{NN}
\eee
The real parameters stand for the magnetic field $h$, the strength $D$ of the DM current and the anisotropy 
$\gamma\in [0,1]$.  
The model is mapped by the Jordan-Wigner transformation 
to the following fermionic one: 
\bee \frac{1}{2}\,H&=&\sum_j (b_j^\dagger\, b_{j+1}\,(1-iD)+b_{j+1}^\dagger \,b_j\,(1+iD)+\nonumber \\
&& +\gamma \,(b_j^\dagger b_{j+1}^\dagger-b_j b_{j+1})
-2h\,b_j^\dagger\,b_j)\ \label{hamig}.
\eee
One can analyse whether the one-particle spectrum determined from (\ref{ops}):
\beq \frac{\Lambda(\theta)}{2}=D \sin\theta+\sqrt{(\cos\theta-h)^2+\gamma^2\sin^2\theta}\eeq
vanishes or not at some $\theta$ to arrive at the following phase diagram:
\begin{center}\includegraphics[width=8cm]{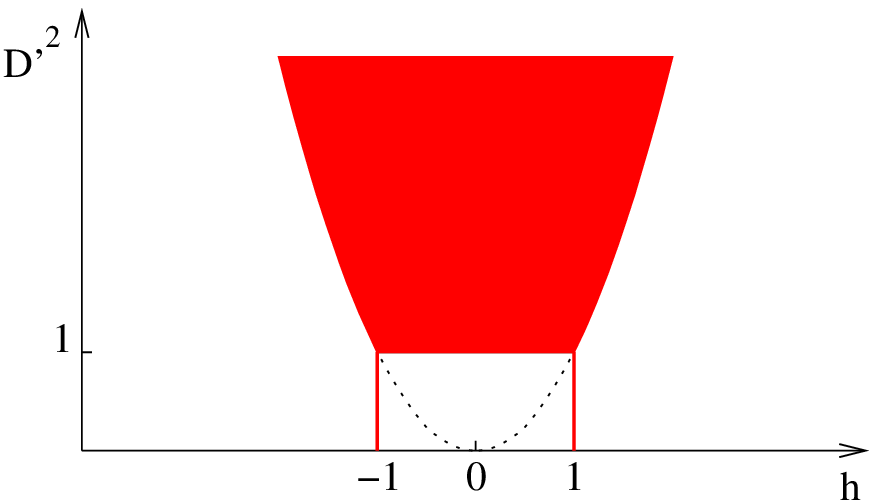}\end{center}
The parameter $D'$ is defined by
\bee 
D'=\sqrt{D^2+1-\gamma^2} \, , \label{Dprime}
\eee
and the critical regions are (i) 
the connected one between the $D'^2=h^2$ paraboles and the $D'^2=1$ line and (ii) the $|h|=1$ line segments. One 
immediately observes from the form of the two point correlations (\ref{symbol}) that the ground state of the non-critical regions are given
by the $XY$ model as $M(\theta)\equiv 0$ for all $\theta\in [0,2\pi)$ and it is via $M(\theta)$ that $\langle m_j m_l\rangle$ depends on 
$D$.  

Let us turn now to the special case of $\gamma=1$ that is, 
the Ising model with DM term. The phase diagram 
is the same
as above with $D'^2=D^2$. The case $h=1$ can be solved by noticing 
that this case belongs to the class of models with $T$ matrix of the genuine Toeplitz type.
We have only two non-zero elements $T_1=1/2,\,T_2=-D/4$, 
which gives $i\,T(\theta)=4\sin \theta(1-D\cos \theta)$ 
for the numerator of the
symbol (\ref{iglo}). The entropy asymptotics for this class is given via 
$S_L=S_{2L}^{A=iT}/2$, as explained after Eq.~(\ref{iglo}), the superscript refers to 
the gauge-invariant model, 
whose Toeplitz symbol is $A$. Its entropy is given 
given by Eq.~(\ref{entnob}). The final result reads
\beq S_L=\left\{\begin{array}{ll}\frac{1}{3}\ln L+\frac{1}{12}\ln \left(1-\frac{1}{D^2}\right)+C_{IsDM} & |D|>1
\\\\
\frac{1}{6}\ln L+C_{Is} & |D|\le 1\end{array}\right.
\eeq
with
\been C_{IsDM}&=&\frac{1}{3}\left(1+\gamma_E+(1-6I_3)\ln 2\right)\approx 0.726067\\
     C_{Is}&=&\frac{1}{6}\left(1+\gamma_E+(2-6I_3)\ln 2\right)\approx 0.478558\eeen
\section{Seeming violations of the Calabrese-Cardy 
formulas \label{Violations}}

In this section we will discuss two "anomalies" 
of the entropy asymptotics, which can appear 
at and in the vicinity of reflection-symmetry-breaking 
critical points and which seemingly do not fit 
the Calabrese-Cardy formulas.
The first concerns the growth of
the saturation entropy as we approach such critical points,
while the second is about the breaking of reflection symmetry
in the finite-size scaling of the entanglement entropy. 
We will discuss how we can interpret these anomalies
to keep the validity of the Calabrese-Cardy formulas.
%in the vicinity of
%a critical point where also reflection symmetry is broken, 
%the second is about the breaking of reflection symmetry
%. These  

\subsection{Anomalous behavior of the saturation entropy}

As mentioned in the Introduction, the formula for the 
saturation value of the entanglement entropy of a block of 
spins near a critical point reads
\bee
S_{sat}=\frac{c}{3} \ln \xi + const \, , \label{sat-2}
\eee
where $c$ is the central charge belonging to the critical point.
We have seen in the previous section that 
considering the region $0 < D' < 1$ (see Eq.~(\ref{Dprime}) 
for the definition of $D'$), the XY
model with DM interaction
is critical when $h= \pm 1$ and the corresponding central
charge is $c_{{\rm XY-DM}}=1$, while for the model without
DM interaction ($D=0$) the central charge of the 
critical line (at $h= \pm 1$) is $c_{{\rm XY}}=\frac{1}{2}$. 
However, we have also shown that when
$h \ne \pm 1$ the ground state does not the depend on
$D$ in the non-critical region $0 \le D' < 1$. Approaching the 
critical $h = \pm 1$ line in this region,
the divergence of the saturation entropy (which is hence independent
of $D$) is consistent with formula (\ref{sat-2}) in case the central 
charge is $c=\frac{1}{2}$, as can be
seen from the results in \cite{itsmezzadrimo,EER}. Hence the 
formula is not valid for the XY model with DM 
interaction, since for that model the central charge is $1$.
This situation is typical for quadratic models with
reflection symmetry breaking: We have seen in Section \ref{Twopoint} that
the ground state of the Hamiltonian Eq.~(\ref{Hqfree}) and the central charge 
do not depend on Im$A_{i,j}$ at a non-critical point, while they may depend on it
at a critical one.

Hence, in order to understand the failure of formula (\ref{sat-2}), 
and to formulate a possible reinterpretation in the case at hand, let us
first look at an other anomaly, which is, in some sense, similar. 
In the $XX$-model with transverse magnetic field
\bee
H=\sum_i \sigma^{x}_i \sigma^{x}_{i+1} +  \sigma^{y}_i 
\sigma^{y}_{i+1} + h \sigma^{z}_i \, , \nonumber
\eee
there is quantum phase transition at the points 
$h=\pm 1$.
In the region $-1 < h <1 $ the ground state of the
model is critical with algebraically decaying truncated correlation 
function 
$C^{xx}(n)=\langle \sigma^x_{i} \sigma^{x}_{i+n} \rangle -C^{xx}_0$ (where 
$C^{xx}_0=\lim_{n\to \infty} \langle \sigma_i^x\sigma_{i+n}^x\rangle$) and 
a diverging entropy asymptotics $S_L=\frac{1}{3} \ln L +k$,
while outside this region the ground state is either the all-spin-up or
all-spin-down state (depending on the sign of $h$). 
Hence approaching the critical region from the non-critical one, 
the saturation entropy
will not diverge, however, this does not contradict formula (\ref{sat-2}), 
since there is no diverging correlation length either. When we enter 
the critical region the state will change suddenly in such a way that
the correlation function $C^{xx}(n)$ that was 
identically zero in the noncritical region
suddenly will be nonzero and even quasi-long-ranged (decay algebraically
with $n$), i.e., an "infinite correlation length" appears instantaneously.
This is, in some sense, a degenerate situation, because considering
a bigger parameter space, e.g., the XY model with transverse magnetic
field
\bee
H=\sum_i (1+\gamma) \sigma^{x}_i \sigma^{x}_{i+1} +  
(1-\gamma)\sigma^{y}_i \sigma^{y}_{i+1} + h \sigma^{z}_i \, , \nonumber
\eee
and approaching the critical line $ (-1 < h < 1, \gamma=0 )$
by fixing the value of $h$ (between $1$ and $-1$) and taking
$\gamma \to 1$, we will observe a diverging correlation length 
and a diverging saturation entropy satisfying formula Eq.~(\ref{sat-2}),
as can be seen from the results in \cite{itsmezzadrimo,EER}.

A similar situation, but in a more complicated form, occurs in the
XY model with DM interactions. At the critical point, the  
correlation functions 
$C^{xx}(n)$ and 
$C^J(n)=\langle\sigma_i^x\sigma_{i+n}^y-\sigma_i^y\sigma_{i+n}^x\rangle$
decay algebraically. However, away from criticality $C^J(n)$
is identically zero, while $C^{xx}(n)$ behaves "in a normal way", i.e., 
it decays exponentially with $n$ and the correlation
length $\xi_s$ length, characterizing the exponential decay, 
diverges as one approaches the 
critical line $h = 1$. We can think that there are two independent 
critical "modes" both with $c=\frac{1}{2}$, one is behaving normally, 
the other in an anomalous way, hence in Eq.~(\ref{sat-2}) we should 
only insert the central charge of the normally behaving mode. 
This picture could be made more convincing and precise, if 
one could show that, similarly to the previously 
mentioned XX case, this anomalous behaviour is 
a degenerate one by considering a bigger subspace, e.g.,
the XYZ chain with magnetic field and DM interaction,
i.e., adding the term  $\sum_i \Delta \sigma^z_i \sigma^z_{i+1}$ 
to the Eq.~(\ref{NN}). We conjecture that 
approaching the mentioned critical point in such a bigger
parameter space the generic behavior 
induces normally diverging correlation lengths for both
$C^{xx}(n)$ and $C^J(n)$, denoted by $\xi_{xx}$ and $\xi_J$, respectively;
and the entropy will scale according to a "generalized" Calabrese-Cardy 
formula of the form of
\bee
S_L=\frac{1}{2} \ln \xi_{xx} + \frac{1}{2} \ln \xi_J + const \, .
\eee  
We have started to study this conjecture numerically,
and the results will be the subject of a forthcoming publication.

\subsection{Breaking of reflection invariance in the finite size scaling
of the entanglement entropy}

The other feature we will investigate is whether
(and to which extent) the breaking of reflection invariance
can be observed as a finite size effect in the 
scaling of the entropy. 
More precisely, consider a finite spin chain with a quadratic Hamiltonian
of length $N$, and compute the entropy $S(L,N)$ of the restriction 
of the ground state to the first $L$ sites. 
The Calabrese-Cardy formula (\ref{finsize}),
which has been confirmed (up to subleading corrections) analytically and numerically for many 
reflection-symmetric models \cite{LSCAe,LSCA,caca}, suggests a reflection-invariant
form $S(L,N)=S(N-L,N)$. 
The question we ask is whether this symmetry of the ground state can be broken
for a quadratic Hamiltonian, which is not invariant and whether the symmetry breaking
survives the limit $N \to \infty$ (with $L/N$ fixed)?

First, we should notice that the reflection invariance of the entropy function
$S(L,N)$ can only be broken if neither the matrix $A$ nor $B$ is real for the following reasons.
We saw that the Hamilton operator (\ref{hamig}) is invariant (and so is the unique ground state) unless 
Im$A\neq 0$. For the case Im$B=0$, one should consider the transformations $b_i\to b_{N-i}$ and 
$b_i\to b_i^\dagger$ and determine the transformed density matrices restricted to the first $L$ sites of the chain.
For the case Im$B=0$ they are identical (both transformations lead to changes $A\to A^\dagger,\,B\to -B$ in the
Hamiltonian). The first one corresponds to the reflection we are interested in, 
whereas the second is a local transformation of the
chain and those preserve the entanglement entropy. 

As noted in Sec.~\ref{Nearest}, the nearest neighbor quasifree Hamiltonians can always 
be transformed by local transformations such that $B$ is real. Hence to
have a symmetry breaking entropy function we have to consider next-to-nearest
neighbour Hamiltonians. The particular Hamiltonian we investigated was 
\bee
&&H=\sum_{i=1}^{N} \Big(
t^{\phantom*}_1 b^{\dagger}_i b^{\phantom\dagger}_{i+1} +
t^{*}_1  b^{\phantom\dagger}_{i+1} b^{\dagger}_i +  
t^{\phantom*}_2  b^{\dagger}_i b^{\phantom\dagger}_{i+2} +
t^{*}_2  b^{\phantom\dagger}_{i+2} b^{\dagger}_i\nonumber \\
&& +  p^{\phantom*}_1  b^{\phantom\dagger}_ib^{\phantom\dagger}_{i+1} -
p^{*}_1  b^{\dagger}_{i}b^{\dagger}_{i+1} +p^{\phantom*}_2 
 b^{\phantom\dagger}_ib^{\phantom\dagger}_{i+2} -
p^{*}_{2} b^{\dagger}_{i}b^{\dagger}_{i+1} 
+ h \, b^{\dagger}_i b^{\phantom\dagger}_i\Big) \, , \nonumber
\eee
with the following parameters: $t_1=7+28i$, $t_2=4+5i$, $p_1=11+10i$,
$p_2=3+4i$, and $h=12$.

\begin{figure}
\includegraphics[width=2.8cm]{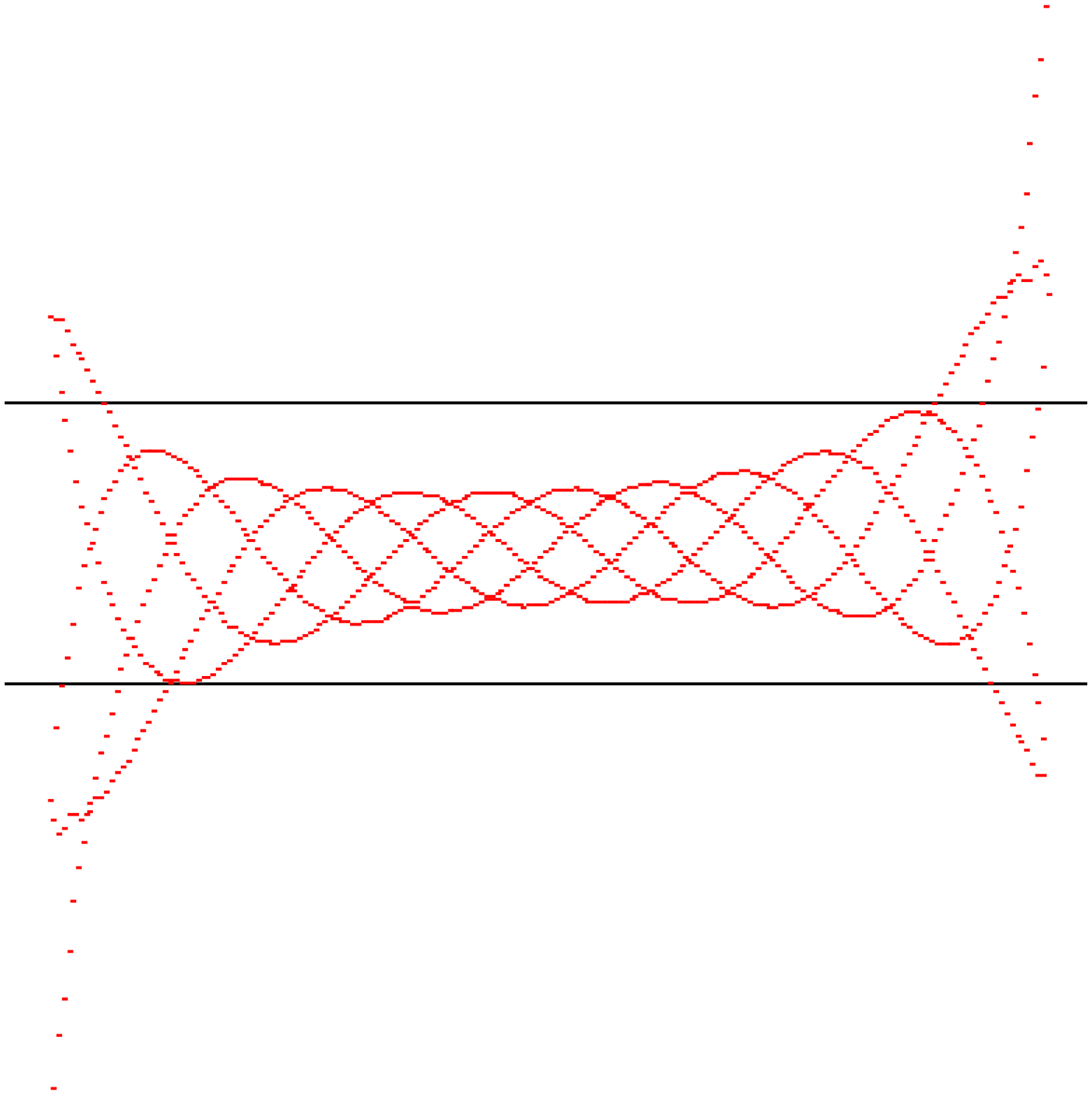}\includegraphics[width=2.8cm]{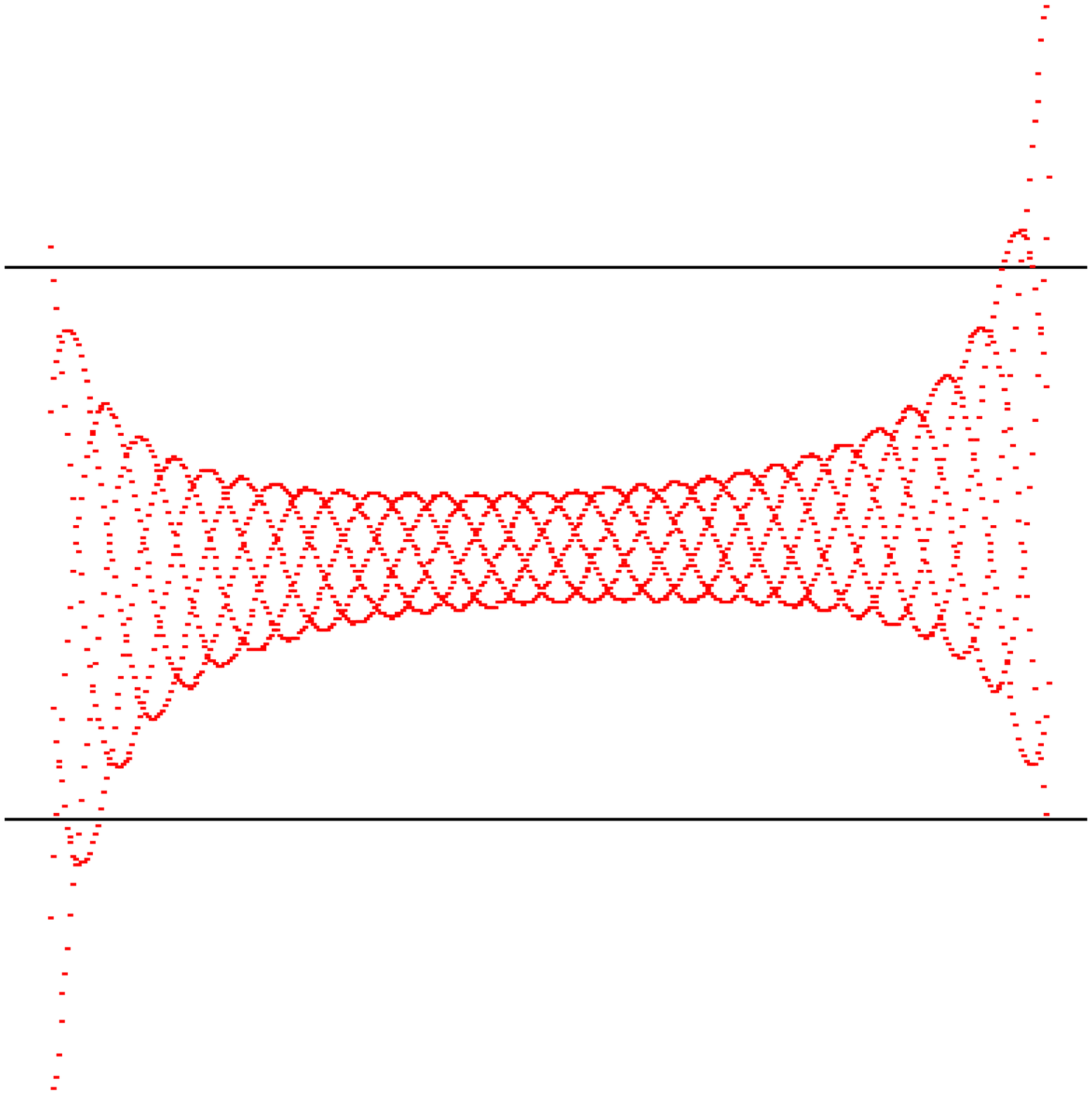}\includegraphics[width=2.8cm]{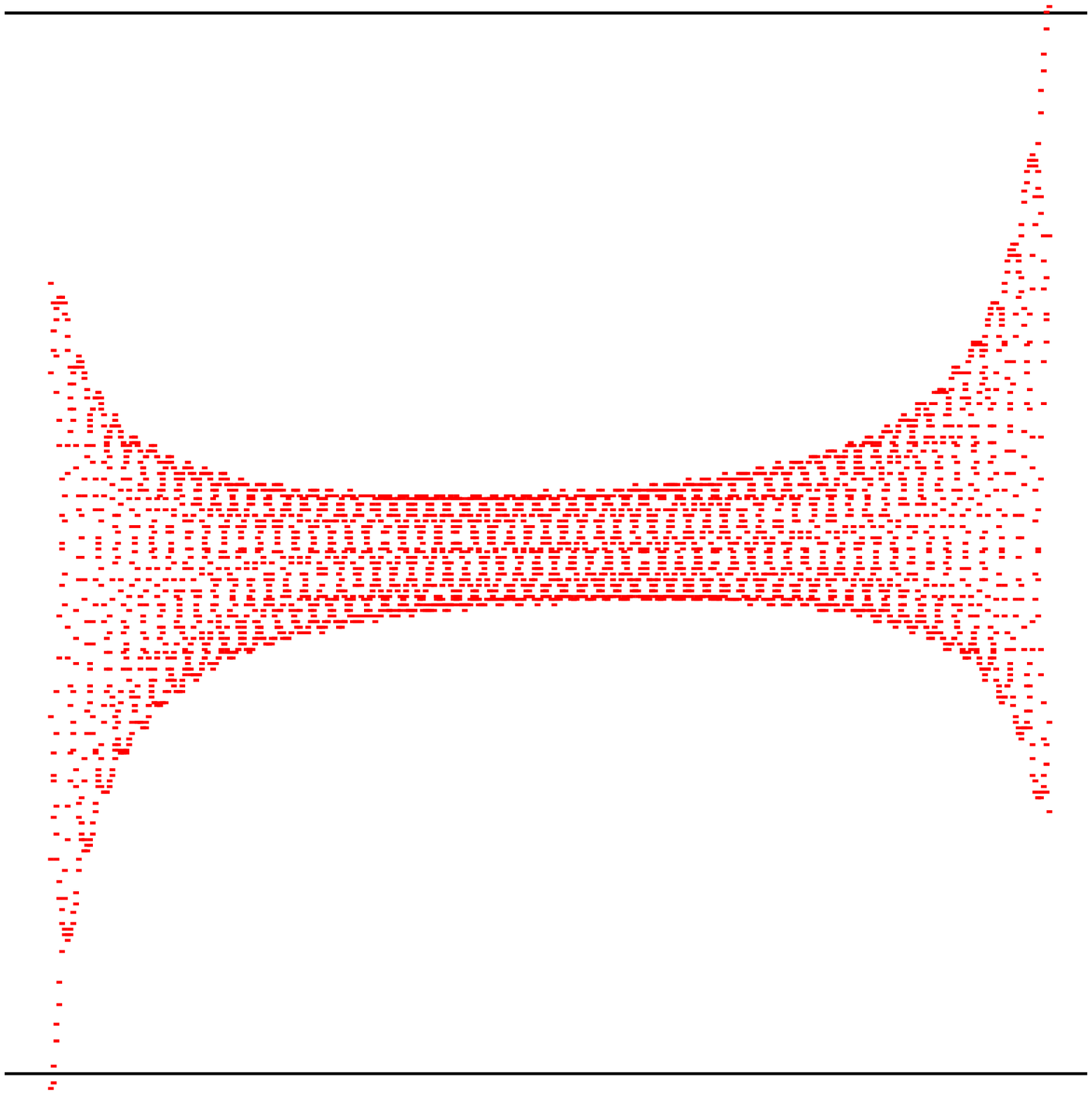}
\caption{\label{breaking} The figures show 
$\Delta S\equiv S(L,N)-S(N-L,N)$, the difference  
between the entropies of the ground state restricted
to the left- and rightmost $L$ modes of the chain with $N=1024, 2048, 4096$. The black lines indicate $\Delta S=\pm 10^{-3}$. 
The plots show, in all three cases, the range $x\equiv L/N\in [0.04,0.96]$. Beside the feature that the 
function $N\mapsto \Delta S(N)$ decreases for any fixed $L$, it also has an additional oscillating structure. One finds that the
analytic function $p\,(N/L+N/(N-L))\cos 2\pi L(\lambda/N-1/5)$ fits this structure rather well with a suitable constant $p$
for the amplitude and the $N-$independent wavelength $\lambda$. The explanation of this behaviour is under investigation.}

\includegraphics[width=6cm]{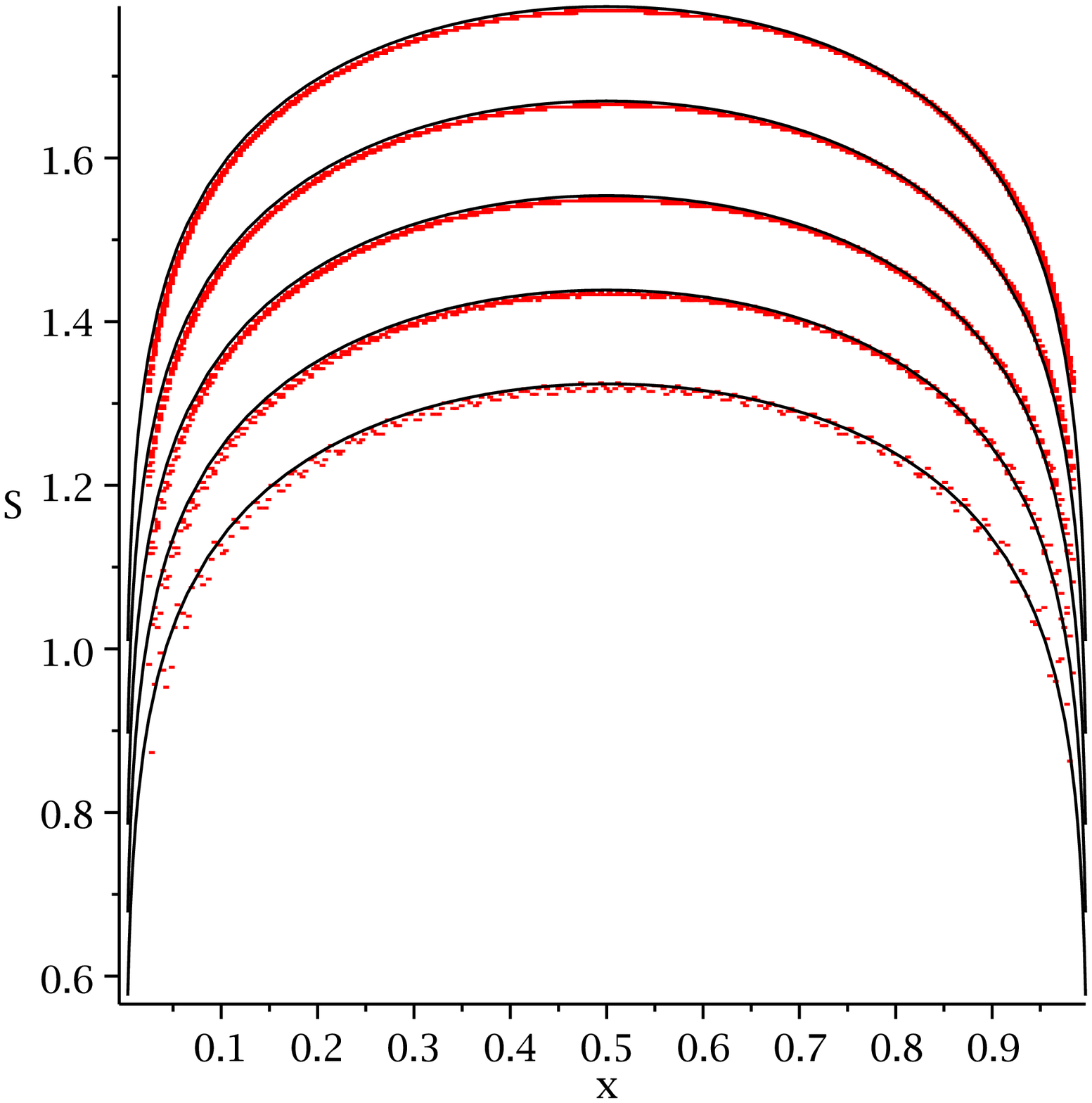}\caption{\label{ccf}Here we can 
see the entropies 
$S(L,N)$ for $N=256, 512, 1024, 2048, 4096$ and the corresponding 
Calabrese-Cardy curves. The central charge is a fit parameter and
it converges to the physical value $c=1$ fast (the deviation 
decreases roughly linearly with $N$ from $0.038$ at $N=256$ to 
$0.003$ at $N=4096$).}
\end{figure}

The numerical results depicted in 
Fig.~\ref{breaking} demonstrate that the reflection symmetry of the
entropy function $S(L,N)$ is indeed broken. However, it is also visible that 
the deviation $S(L,N)-S(N-L,N)$ goes to zero in the limit $N \to \infty$ for any fixed $L$. 
Moreover, we can see in Fig.~\ref{ccf} that in this limit the curves nicely converge to the Calabrese-Cardy formula.
Hence we conclude that, according to our numerical results, the
reflection symmetry of the entropy function can be broken, but 
its scaling limit shows no such breaking, and the Calabrese-Cardy formula 
is valid.

\section{Summary and overview}

In this paper we studied the entanglement entropy asymptotics of
spin chains that can be mapped to quasifree
fermionic models given by the sum of a 
gauge-invariant term (parametrized by 
a selfadjoint matrix $A$) and a non-gauge-invariant one (parametrized by an 
antisymmetric matrix $B$). Many models of physical importance 
belong to the class of complex $A$ (and $B$), implying the breaking of
reflection symmetry). The entanglement 
properties of these systems have hardly been 
addressed in the literature before, 
hence we concentrated on these cases.

We have determined the two-point functions of the majorana 
operators in complete generality. A novelty following from 
this investigation is that the ground state can only be 
reflection symmetry breaking if it is critical.

We have been able to write down the  
analytic expression of the entropy asymptotics 
for the most general gauge-invariant models,
and also extended these results for
certain non-gauge-invariant models.
A detailed investigation of the nearest neighbor case 
was carried out. We have derived the explicit form of
the entanglement entropy asymptotics for 
the Ising model with Dzyaloshinskii-Moriya 
interaction at the critical point, which was unknown until now. In the 
noncritical regime, we demonstrated that 
the ground state is independent of 
the DM coupling, thus the 
entropy asymptotics given in \cite{itsmezzadrimo} without 
the DM term is valid also here. This indicated violations 
of the formula for the saturation entropy $\sim c\log \xi$
near the critical point $|h|=1$. We have given a
possible "physical explanation" for this.

Concerning the general landscape of the block entropy asymptotics 
of quasifree models, we extended the general
knowledge to a large extent, nevertheless, the general case 
remains to be a difficult unsolved mathematical 
problem \footnote{Assuming that three of the 
four polynomials $A^s(z), A^a(z), B^s(z), B^a(z)$ are linearly 
independent, the entropy is unknown.}. Even when specifying the discussion 
to the nearest neighbor case, there remains a 
surprisingly large region of the critical regime, for which the 
scaling of the block entropy still remains an open problem.

Finally, we carried out numerical checks for the investigation
of finite size effects. We used a model Hamiltonian with 
next-to-nearest neighbor interaction, which exhibited reflection 
symmetry breaking in the finite-size-scaling of the entanglement entropy. 
The deviation was demonstrated to converge to zero quickly by increasing 
the size of the chain, while the block entropy converged to the 
asymptotic Calabrese-Cardy formula.   

\subsection*{Acknowledgements}
We thank Lorenzo Campos Venuti for discussions and Alexander R. Its for a correspondence. The work was supported 
by EU-STREP Project COQUIT (grant no. 233747).
%%%%%%%%%%%%%%%%%%%%%%%%%%%%%%%%%%%%%%%%%%%%%%%%%%%%%%%%%%%%%%%%%%%%%%%%%%%%%%%%%%%%%%%%%%%%%%%%%%

\end{document}